%% file: sample-manuscript.tex
\PassOptionsToPackage{table,xcdraw}{xcolor}
\documentclass[manuscript]{acmart}

\usepackage{booktabs} 
\usepackage{multirow}
\usepackage{tabularx}
\usepackage{fancybox}
\usepackage{graphicx}

\def\BibTeX{{\rm B\kern-.05em{\sc i\kern-.025em b}\kern-.08emT\kern-.1667em\lower.7ex\hbox{E}\kern-.125emX}}
    
\copyrightyear{2021}
\acmYear{2021}
\setcopyright{acmlicensed}\acmConference[IUI '21]{26th International Conference on Intelligent User Interfaces}{April 14--17, 2021}{College Station, TX, USA}
\acmBooktitle{26th International Conference on Intelligent User Interfaces (IUI '21), April 14--17, 2021, College Station, TX, USA}
\acmPrice{15.00}
\acmDOI{10.1145/3397481.3450694}
\acmISBN{978-1-4503-8017-1/21/04}

\definecolor{highlight}{HTML}{EFEFEF}
\definecolor{rowcol}{rgb}{0.9,0.9,0.9}
\newcommand{\rot}[1]{\begin{tabular}{@{}c@{}}\rotatebox[origin=c]{90}{#1}\end{tabular}}
\newcommand{\mytabhead}[1]{\textbf{\normalsize #1}}
\newcommand{\mytabheads}[1]{{\small #1}}
\newcommand{\mytabheadxs}[1]{{\footnotesize #1}}
\newcommand{\newexample}{\vspace{1.25mm}\linebreak}
\newcommand{\sysprop}{system qualities}

\newcommand{\knowprocess}{process knowledge}
\newcommand{\knowinteract}{interaction knowledge}
\newcommand{\knowmeta}{meta knowledge}

\definecolor{highlightboxColour}{rgb}{0.95,0.95,0.95}

\begin{document}
\title{How to Support Users in Understanding Intelligent Systems? Structuring the Discussion}

\author{Malin Eiband}
\email{malin.eiband@ifi.lmu.de}
\affiliation{%
  \institution{LMU Munich}
  \streetaddress{Frauenlobstra{\ss}e 7a}
  \city{Munich}
  \state{Germany}
  \postcode{80337}
}

\author{Daniel Buschek}
\email{daniel.buschek@uni-bayreuth.de}
\affiliation{%
  \institution{Research Group HCI + AI, Department of Computer Science, University of Bayreuth}
  \streetaddress{Universit\"atsstra{\ss}e 30}
  \city{Bayreuth}
  \postcode{95447}
  \state{Germany}
}

\author{Heinrich Hussmann}
\email{hussmann@ifi.lmu.de}
\affiliation{%
  \institution{LMU Munich}
  \streetaddress{Frauenlobstra{\ss}e 7a}
  \city{Munich}
  \state{Germany}
  \postcode{80337}
}

\renewcommand{\shortauthors}{Eiband, Buschek and Hussmann}

\input{sections/00-abstract.tex}



\maketitle

\input{sections/01-introduction.tex}

\input{sections/01a-contribution.tex}

\input{sections/02-scope.tex}

\input{sections/03a-mindsets.tex}
\input{sections/03b-involvement.tex}
\input{sections/03c-outcomes.tex}

\input{sections/04-discussion.tex}

\input{sections/05-conclusion.tex}

\begin{acks}
This project is funded by the Bavarian State Ministry of Science and the Arts and coordinated by the Bavarian Research Institute for Digital Transformation (bidt).
\end{acks}

\bibliographystyle{ACM-Reference-Format}
\bibliography{bibliography.bib}

\end{document}

%% file: sections/00-abstract.tex
\begin{abstract}
The opaque nature of many intelligent systems violates established usability principles and thus presents a challenge for human-computer interaction. Research in the field therefore highlights the need for transparency, scrutability, intelligibility, interpretability and explainability, among others. While all of these terms carry a vision of supporting users in understanding intelligent systems, the underlying notions and assumptions about users and their interaction with the system often remain unclear.

We review the literature in HCI through the lens of implied user questions to synthesise a conceptual framework integrating user mindsets, user involvement, and knowledge outcomes to reveal, differentiate and classify current notions in prior work. This framework aims to resolve conceptual ambiguity in the field and enables researchers to clarify their assumptions and become aware of those made in prior work. We thus hope to advance and structure the dialogue in the HCI research community on supporting users in understanding intelligent systems.

\textbf{To appear in the Proceedings of the 26th ACM Annual Conference on Intelligent User Interfaces (IUI '21), peer reviewed author version 5.0, 18. February 2021}

\end{abstract}

%% file: sections/01-introduction.tex
\section{Introduction}

Interactive intelligent systems violate core interface design principles such as predictable output and easy error correction~\cite{Amershi2014, Dudley2018}.
This makes them hard to design, understand, and use~\cite{QYang2020} -- an observation that has already been made decades earlier~\cite{Hook2000}, but it is only in the last years that machine learning has increasingly penetrated everyday applications and thus refuelled the discussion on how we want interaction with such systems to be shaped. 

One particularly challenging property of intelligent systems is their opaqueness. As a result, researchers~\cite{Alkhatib2019, ACM2017, Hager2017}, practitioners~\cite{Chander2018}, policy-makers~\cite{GDPR} and the general public~\cite{Kuang2017web} increasingly call for intelligent systems to be transparent~\cite{Eiband2018}, scrutable~\cite{Kay2013}, explainable~\cite{Rader2018}, intelligibile~\cite{Lim2010a} and interactive~\cite{Dudley2018}, among others, which we will henceforth refer to as \textit{system qualities}.
Work on the \sysprop{} follows a joint and urgent maxim: Designing interaction in a way that supports users in understanding and dealing with intelligent systems despite their often complex and black-box nature.

Linked by this shared goal, the diverse terms are often employed interchangeably -- and yet, prior work implies divergent assumptions about how users may best be supported.
For instance, work on interpretability has recently been criticised for unclear use of the term~\cite{Doshi-Velez2017, Lipton2018}, a survey on explainability in recommenders found incompatible existing taxonomies~\cite{Nunes2017}, and discussions about system transparency and accountability revealed diverging assumptions (i.e. disclosing source code vs system auditing through experts)~\cite{Edwards2017}. A recent HCI survey shows the fractured terminological landscape in the field~\cite{Abdul2018}.

In particular, for supporting user understanding of intelligent systems, clarifying concepts and connecting diverse approaches is crucial to advance scholarship, as pointed out in a recent ``roadmap'' towards a rigorous science of interpretability~\cite{Doshi-Velez2017}. More generally speaking, a lack of conceptual clarity impedes scientific thinking~\cite{Hornbaek2017} and presents challenging problems for researchers in the respective field:
First, a lack of overarching conceptual frameworks renders new ideas difficult to develop and discuss in a structured way.
Second, blurred terminological boundaries impede awareness of existing work, for example through varying use of keywords. 
Third, new prototypes often remain disconnected from the existing body of design solutions. 
To address this, we need a clearer conceptual understanding of the \textit{assumptions} that underlie how prior work envisions to foster user understanding of intelligent systems. 

\begin{figure*}[t]
    \centering
    \includegraphics[width=0.9\textwidth]{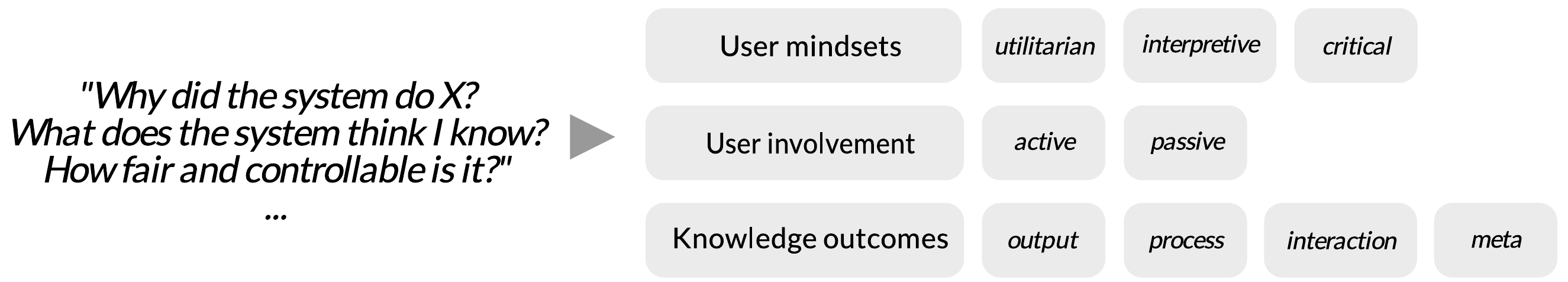}
     \caption{Our framework for structuring the discussion of how to support users in understanding intelligent
    systems: We examine user questions in the literature (left, examples) to synthesise three categories overarching prior work (centre), namely assumed \textit{user mindsets}, \textit{user involvement} and \textit{knowledge outcomes}. We discuss divergent instances of each category (right) to differentiate approaches and solution principles in the literature.}
     \label{fig:teaser}
\end{figure*}

In this paper, we thus aim to answer the following research questions:

\begin{quote}
    \textit{\textbf{RQ1:} Which assumptions about users and interaction with intelligent systems do researchers make when referring to the \sysprop?}
\end{quote}
\begin{quote}
    \textit{\textbf{RQ2:} How can we structure and differentiate these assumptions?}
\end{quote}

%% file: sections/01a-contribution.tex
\section{Contribution and Summary}

We analyse both theoretical concepts and prototype solutions through the lens of implied \textit{user questions} and synthesise a conceptual framework integrating \textit{user mindsets}, \textit{user involvement} and \textit{knowledge outcomes} to reveal, differentiate and classify notions of supporting user understanding of intelligent systems in prior work.

Our analysis revealed three categories that capture and differentiate current assumptions about users and interaction with intelligent systems from an HCI perspective (also see Figure~\ref{fig:teaser}):
\begin{enumerate}
    \item \textit{User mindsets} -- what users seek to know (e.g. do they want to know \textit{why the system did X}, \textit{what it was developed for}, \textit{how trustworthy it is}, etc.),
    \item \textit{User involvement} -- how users gain knowledge (e.g. do they actively inquire into the system or do they get presented information by it), and
    \item \textit{Knowledge outcomes} -- what kind of knowledge users gain (e.g. about a specific output or how to correct errors).
\end{enumerate}

In particular, as we will describe later in more detail, we argue that these three categories are linked to users' intentions when using a system, influence the direction of information transfer between user and system, and reflect the envisioned outcome of the \sysprop, respectively.

Our view helps to resolve conceptual ambiguity in the field and provides researchers with a framework to clarify their assumptions and become aware of those made in prior work. We thus hope to advance and structure the dialogue in the HCI research community on supporting users in understanding intelligent systems.

%% file: sections/02-scope.tex
\section{Scope and Foundations}
Before we present the results of our analysis in detail, we first discuss fundamental conceptual prerequisites for our work and locate our own perspective.
\subsection{Intelligent Systems}
Our work focuses on interaction with \textit{intelligent} systems.
Following Singh~\cite{Singh1994}, a system is intelligent if we need to ``attribute cognitive concepts such as intentions and beliefs to it in order to characterize, understand, analyze, or predict its behavior''. 
While we are aware of the fact that many other definitions of intelligent systems exist, Singh's definition hints at the potential complexity of intelligent systems and the resulting challenges for \textit{users} to \textit{understand} them, and thus motivates work on supporting users in doing so, including this paper.

\subsection{A Pragmatic View on Supporting User Understanding}

One can identify two opposing perspectives in the larger discussion of supporting users in understanding intelligent systems: A normative and a pragmatic one~\cite{Eiband2018a}. 

The normative perspective is visible in the ethical discourse about intelligent systems or reflected in legislation. It provides users with what has been called a ``right to explanation''~\cite{Goodman2016}, such as recently articulated in the GDPR~\cite{GDPR}, and ties lawful use of intelligent systems to the ability to make users understand their decision-making process~\cite{Hildebrandt2016}. While highly valuable for a legal basis for interaction with intelligent systems, this perspective stems from ethical and moral reflection, not from user needs for interaction with a concrete system. As such, it often lacks specifics on how to implement its claims in a way that benefits users in practice. 

In this paper, we therefore adopt the \textit{pragmatic} perspective, which strives to best support users during interaction with intelligent systems. We define the purpose of this support by transferring a statement by Lynham from philosophy of science~\cite{Lynham2002} to the context of our work: \textit{Supporting users in understanding intelligent systems means helping people to use a system better and in more informed ways, and to better ends and outcomes.}

We argue that this perspective captures well and articulates a core assumption of work on the \sysprop: We as HCI researchers in the field of intelligent systems strive to create interfaces and interactions that are explainable, understandable, scrutable, transparent, accountable, intelligible, and so on, precisely because we envision users to then interact with these systems in more informed, effective and efficient ways.

\subsection{User Knowledge and Understanding}

In general, HCI has widely adopted mental models~\cite{Johnson-Laird1989} as representations of the knowledge users possess about a system~\cite{Norman2013}, and this is no different in work on the \sysprop{} (e.g.~\cite{Eiband2018a, Kulesza2012, Tullio2007}). Mental models originate from a constructivist perspective on knowledge, where knowledge is seen as individually constructed, subjective interpretations of the world, based on previous experiences and assumptions~\cite{VonGlasersfeld1989}. In this paper, we adopt this perspective, and use \textit{knowledge} interchangeably with \textit{understanding}. 
Moreover, we assume that knowledge is gained through the \textit{transmission of information} between user and system.

%% file: sections/03a-mindsets.tex
\begin{table*}
\centering
\scriptsize
\newcolumntype{L}{>{\raggedright\arraybackslash}X}
\renewcommand{\arraystretch}{1.5}
\resizebox{0.95\textwidth}{!}{
\begin{tabularx}{\textwidth}{@{}>{\raggedright\arraybackslash}p{5.5em}>{\raggedright\arraybackslash}p{20em}>{\raggedright\arraybackslash}p{7em}>{\raggedright\arraybackslash}p{10em}L@{}}
\toprule
\mytabheadxs{\textbf{System quality}} & \mytabheadxs{\textbf{User questions (examples)}} & \mytabheadxs{\textbf{User mindsets}} & \mytabheadxs{\textbf{User involvement}} & \mytabheadxs{\textbf{Knowledge outcomes and qualities}} \\ 
\midrule
Accountability & How fair and controllable is the system?~\cite{Rader2018} How fair are system decisions?~\cite{Binns2018} How will algorithmic-decision making impact my (social) work practices?~\cite{Brown2019} How can I as a designer/developer of decision systems support human values?~\cite{Holstein2019, Veale2018} & Users \textit{critique} the system, often in a wider context beyond use (e.g. legal, ethical concerns). & Users get informed by the system, or by a third party reporting on the system (e.g. journalists)~\cite{Diakopoulos2015, Diakopoulos2017}.
People discuss the system and/or challenge its implications beyond use~\cite{Schlesinger2018}, e.g. in its organisational context~\cite{Brown2019, Holstein2019, Veale2018}. & Users seek reflection on \textit{outputs}, \textit{processes}, as well as on reasons behind and implications of the system (\textit{meta}). What is relevant -- and to what extent -- may depend on the context of the system's deployment. \\

\cellcolor[HTML]{EFEFEF} Debuggability (end-user debugging) & \cellcolor[HTML]{EFEFEF} How can I correct system errors? 
How can I tell the system why it was wrong?~\cite{Kulesza2011} 
How can I give effective feedback to the system? 
Which objects do I have to change? 
How do changes affect the rest of the system? 
What does the system's feedback mean? 
How can I detect a system error? 
How can I find out the cause for an error? 
How can I solve this error?~\cite{Kulesza2008} & \cellcolor[HTML]{EFEFEF} Users gain insight into the system to \textit{fix its errors}. & \cellcolor[HTML]{EFEFEF} Users fix the system's errors. & \cellcolor[HTML]{EFEFEF} Users need to understand \textit{outputs}, \textit{processes}, and \textit{interactions} to give good feedback and correct the system. Users make the system more relevant to them by correcting system errors. \\

Explainability & Can I trust this model?~\cite{Ribeiro2016} 
Should I trust this prediction?~\cite{Herlocker2000, Ribeiro2016} 
What are the strengths and limitations of the system? 
How can I add my knowledge and skills to the decision process?~\cite{Herlocker2000} 
How are input and output related?~\cite{Johnson1993} 
Why does the system think that I want/need X?~\cite{Billsus1999, Donkers2020} Why is this recommendation ranked at the top?~\cite{Tsai2019b} & Users gain insight into the system to \textit{better use} it. & Users get informed by the system. & Users get information about \textit{outputs} and \textit{processes}. Explanations should be relevant to the user.
They should be ``sound and complete'', but not overwhelming~\cite{Kulesza2013, Kulesza2015}.  \\

\cellcolor[HTML]{EFEFEF} Intelligibility & \cellcolor[HTML]{EFEFEF}Why did the system do X? 
How / under what conditions does it do Y? 
Why did not do Y? 
What (else) is it doing? 
What if there is a change in conditions, what would happen?~\cite{Lim2009, Coppers2019} &\cellcolor[HTML]{EFEFEF} Users want to use the system in \textit{better} ways or to gain trust~\cite{Lim2009a, Yang2020}. & \cellcolor[HTML]{EFEFEF}Users actively inquire into the system's inner workings. & \cellcolor[HTML]{EFEFEF}Users seek information about \textit{outputs} and \textit{processes}. Users' demand informs what is relevant. Factors related to system and context influence this~\cite{Lim2009}. \\

Interactivity (interactive machine learning) & How I can assess the state of the learned concept? 
Where does the model fail? 
Why did the system fail in this specific instance?~\cite{Dudley2018} 
How well does the system know the domain? 
How sure is the system that a given output is correct? 
Did the system do a simple or complex thing to arrive at the output?~\cite{Sarkar2015} 
How to combine models?~\cite{Talbot2009} 
Which model works best?~\cite{Amershi2015} & Users inspect the system state to \textit{refine} it or guide its training~\cite{Dudley2018}. & Users iteratively refine the system and guide its training by giving feedback~\cite{Dudley2018}. & Users need to understand \textit{outputs}, \textit{processes}, and \textit{interactions} to guide the system.
What is relevant to know is defined by the machine learning task that users and system solve together.  \\

\cellcolor[HTML]{EFEFEF} Interpretability & \cellcolor[HTML]{EFEFEF}How sensible -- and not arbitrary or random -- is the system?~\cite{Alqaraawi2020, Rader2018}
\textit{Why?} questions~\cite{Gilpin2018}
Can you trust your model?
What else can it tell you about the world?~\cite{Lipton2018} & \cellcolor[HTML]{EFEFEF}Users gain \textit{utilitarian} and \textit{interpretative} insight into the system to bridge the gap between the system's criteria and full real-world context~\cite{Doshi-Velez2017, Kim2015}. &\cellcolor[HTML]{EFEFEF} Users get  information about the system's inner workings~\cite{Doshi-Velez2017, Lipton2018}. &\cellcolor[HTML]{EFEFEF}  Users can access information about \textit{outputs} and \textit{processes}, which may include low-level (expert) information (e.g. on inner states~\cite{Kim2015}). What is relevant to know depends on the user's task with the system. Explicit call for rigorous evaluation~\cite{Doshi-Velez2017}. \\

Scrutability & Why/How did the system do X? 
What else does the system think I (don't) know? 
What would the system do if I did Y? 
What does the system do for other people? 
How can I tell the system what I (don't) want?~\cite{Kay2013} How can I efficiently improve recommendations?~\cite{Balog2019} & Users want to be able to \textit{interpret} the system's decisions.
They may analyse and control it for more \textit{efficient} use. & System decision and behaviour is based on a user model, which users can adequately access and control.
Users make ``real effort''~\cite{Kay2013}. & Users gain understanding of \textit{outputs} and \textit{processes}.
They may also learn about \textit{interactions} to influence how the system uses the user model. Information should be relevant to users, yet they may also learn about what the system considers relevant. \\

\cellcolor[HTML]{EFEFEF} Transparency & \cellcolor[HTML]{EFEFEF}How does the system produce an output (i.e. data sources, reasoning steps)? 
Why did the system do sth. (i.e. justification, motivation behind the system)? 
What is informed by the intelligent system (i.e. reveal existence of intelligent processing)? 
How was the system developed and how is it continually being improved?~\cite{Rader2018} How did the system produce the model?~\cite{Drozdal2020} & \cellcolor[HTML]{EFEFEF}Users \textit{interpret} the system's output and question the underlying mechanisms. & \cellcolor[HTML]{EFEFEF}Users get informed by the system. & \cellcolor[HTML]{EFEFEF}Users seek understanding of \textit{outputs} and \textit{processes}, also beyond use (\textit{meta}). What is relevant -- and to what extent -- may depend on the context of the user's inquiry. \\ \bottomrule
\end{tabularx}
}
\caption{The \sysprop{} in focus of our work through the lens of our framework, along with example user questions and implied user mindsets, user involvement, and knowledge outcomes. Note that we do not attempt to redefine the terms here but rather provide a guiding overview of our coding, which includes some overlap between terms, as found in the literature.}

\label{tab:system_qualities_questions}
\end{table*}

\section{Method}
Here, we shortly describe the process of paper collection and interpretation through which we derived our framework.

\subsection{Theoretical Sampling}\label{sec:sampling}
Our paper set was collected using \textit{theoretical sampling}, an approach to collection of qualitative data introduced by Glaser and Strauss as a part of Grounded Theory~\cite{glaserstrauss1967}. In contrast to statistical sampling, where the sample size is fixed a priori, theoretical sampling gradually defines the sample during the interpretation process until \textit{theoretical saturation} is reached (i.e. the point where further data and interpretation does not further enrich the emerging categories).

We started our sampling by looking for a set of the most widely-adopted \sysprop{} in the field. We did so first through collecting search terms for system qualities based on our experiences as researchers
working at the intersection of HCI and AI for several years and then expanded our set of \sysprop{} through the topic networks presented by Abdul et al.~\cite{Abdul2018} in their 2018 survey of over 12,000 papers at this intersection. Their analysis surfaced many \sysprop{} that we sought to address a priori (e.g. interpretability, scrutability,
explainability), but also related topics (e.g. accountability and different types of transparency).  
With keyword searches on the ACM Digital Library, we then iteratively collected papers on these terms, following the above sampling method. We started out with the most cited papers in the field, which we interpreted as described in Section~\ref{sec:coding} to create first categories. Papers were selected to represent the diversity of approaches in the field, but also according to the below criteria:

\begin{enumerate}
	\item The presented contribution focuses on the \sysprop{} and is linked to \textit{intelligent} systems, and
	\item the contribution involves an HCI perspective (e.g. via a prototype,
	user study, design guidelines, etc.).
\end{enumerate}

Papers were re-visited if necessary, and we integrated new data through snowball searches as well as through updated keyword searches when we extended our \sysprop{} set. Overall, 222 papers contributed to this process before we considered our categories to be theoretically saturated and had defined our framework dimensions. The final set of \sysprop{} emerging from this process included \textit{scrutability, interpretability, transparency, explainability, intelligibility, interactivity (interactive Machine Learning), debuggability (end-user debugging), and accountability}.

The paper set used is available on the project website:
\url{https://www.medien.ifi.lmu.de/howtosupport/}

\subsection{Paper Coding and Interpretation}\label{sec:coding}
We followed the coding process as suggested by Glaser and Strauss~\cite{glaserstrauss1967} and Strauss and Cobin~\cite{strausscobin1990} to code the papers. The coding was done by the first two authors, in joint work on the texts and discussion. 

\subsubsection{Open Coding: Opening Up the Text}
To keep the coding process manageable, we mainly focused on the motivation, contribution and conclusion of a paper, making use of the flexibility of the approach to include other sections as needed for a clearer understanding. 
We first collected open codes about the assumptions on supporting user understanding made in the text. Open coding was guided by so-called basic questions~\cite{glaserstrauss1967} (e.g. \textit{Who?, What?, How?}, etc.).

This step resulted in a first set of (sub-)categories, namely
\textit{Motivation (of the work), Key Phenomena and Constructs, Support as Property or Process, Reference to Specific System Part, Main Goal of Support, Main Challenge of Support, User Role, System Role, Interaction Design Implications}, and \textit{Concrete Realisation}.

\subsubsection{Axial Coding: Eliciting User Questions}
We then created axial codes to refine and differentiate our categories. During this process, we realised that suitable axial codes could be expressed in the form of questions users have about the workings of intelligent systems (e.g. \textit{Why did the system do X?}).
This is in line with prior work: First established by Lim and Dey~\cite{Lim2009,Lim2009a} as a manifestation for users' information demand, such \textit{user questions} have gained popularity in related work on the \sysprop{} as a way of
anchoring design suggestions and solutions, even if these questions are not always elicited
from actual users (e.g., cf. work by Kay and
Kummerfeld~\cite{Kay2013} (scrutability), Kulesza et al.~\cite{Kulesza2011}
(end-user debugging), or Rader et al.~\cite{Rader2018}
(transparency/accountability)).

In our case, coding these questions helped us to extract and contrast underlying perspectives: On the one hand,
they connected prior work on the \sysprop, on the other they revealed conceptual differences, and thus refined our preliminary categories.
During this process, we kept close to the text of a paper. For example, ``Accountability [...] the extent to which participants think the system is fair and they can control the outputs the system produced''~\cite{Rader2018} yielded ``How fair and controllable is the system?'' in our question set. We continuously moved back and forth between the texts and our preliminary categories when integrating new user questions to test our categories against the text passages.

As the outcome of this coding step, we (1) refined the categories elicited in the open coding, and (2) discovered differences and commonalities between the \sysprop{} by grouping user questions according to the quality in focus of a paper (see Table~\ref{tab:system_qualities_questions} for an excerpt; please note that one question could be relevant for multiple \sysprop).

\subsubsection{Selective Coding: Building the Framework}
In the last step, we identified core concepts overarching our categories. Practically, we wrote our user questions on cards we re-arranged on a large wall, plus example papers and stickers for possible core concepts. This set-up was then discussed extensively over several weeks and extended through further paper collection and reviewing. As a result, we identified and defined our framework dimensions presented in Section~\ref{sec:framework}.

\subsection{Limitations}
Our framework should be understood and applied with several limitations in mind:

First, as work on interactive intelligent systems has evolved into a rapidly developing field in the last decade, future approaches might uncover solutions and principles that the presented framework does not cover in its current state. 
Second, this work focuses on eight system qualities, listed in Section~\ref{sec:sampling}. Since this set was shaped through our sampling process, we are confident that it covers a representative part of the diverse approaches and reflections in the field. However, it is not comprehensive. For a detailed overview of possible system qualities, see \citet{Abdul2018}.
Third, our paper set is mainly based on work accessible through the ACM Digital Library. We also included work published through IEEE, Springer, and Elsevier, among others, but only through snowball searches, not as a main database, to keep our paper set at a manually manageable size. 
Overall, these limitations might create the need to build on, extend and adapt the framework in future work.

Finally, this work focuses on HCI \textit{researchers} as a target audience. In Section~\ref{sec:inspiring}, we suggest ideas on how our framework might be applied to commercial systems and thus serve practitioners, too. However, these ideas still need to be validated in the future, for example, through interviews with designers and developers, or by prototyping applications using the framework dimensions and evaluating them with end-users.

\section{A User-Centered Framework for Supporting Users in Understanding Intelligent Systems}\label{sec:framework}

Our analysis revealed three categories that capture and differentiate current
assumptions about users and interaction with intelligent systems. They thus
serve as a conceptual framework for structuring the discussion about supporting
users in understanding intelligent systems from an HCI perspective, as presented in the next sections.

\subsection{What do Users Seek to Know? User Mindsets}\label{sec:mindsets}

The user questions addressed in work on different \sysprop{} imply different
assumptions about \textit{what users seek to know}: For example, searching for
information about \textit{why} a certain system output came into
being~\cite{Lim2009a} implies a very different kind of interest than wanting to
know \textit{how well a program knows a given domain}~\cite{Sarkar2015} or
\textit{how a system was developed and continually improved}~\cite{Rader2018}.
Likewise, this is true for users asking \textit{how to correct system
	errors}~\cite{Kulesza2011} compared to users that want to be merely
\textit{informed} about the presence of algorithmic
decision-making~\cite{Rader2018}. 

To capture these differences, we introduce the category \textit{user mindsets}.
In psychological research, mindsets describe the ``cognitive orientation'' of
people that precede the formation of intentions and planning of successive
actions towards reaching a goal~\cite{Gollwitzer1993}. In the same manner, for
this work we define \textit{user mindsets} as \textit{users' cognitive
	orientation that guides concrete intentions to interact with an intelligent
	system}.
From our analysis emerged three such mindsets that we find in prior work on the
\sysprop: \textit{utilitarian}, \textit{interpretive}, and \textit{critical},
described in detail next. 

\subsubsection{Utilitarian Mindset}
A utilitarian mindset \textit{aims to predict and control system behaviour to
	reach a practical goal}. This mindset carries a strong notion of
\textit{utility} and/or \textit{usability}. 

Consequently, a utilitarian mindset is reflected by many examples in work on
\sysprop{} that imply a very practical view on user inquiry, such as in work
on explainability and intelligibility. For example, users might want to
understand system recommendations to \textit{better compare and find} products they are interested in (\textit{Why was this recommended to me?})~\cite{Pu2006}. Moreover,
users might want to \textit{train more effectively} with an intelligent
fitness coach~\cite{Eiband2018a}, or understand a system ranking they \textit{financially
depend on}, as observed in AirBnB~\cite{Jhaver2018} or social
media~\cite{Bucher2017}. In another example, users worked more efficiently with a system
feedforward based on \textit{What if?} questions~\cite{Coppers2019}. Similarly, user questions in
work on scrutability such as \textit{How can I efficiently improve recommendations?}~\cite{Balog2019} imply a utilitarian mindset.

On a meta-level, this mindset can also be found in work on
interactive machine learning and end-user debugging. Research in these areas
addresses user questions such as \textit{How can I assess the state of the
	learned concept? Where does the model fail?}~\cite{Dudley2018}, \textit{How sure
	is the system that a given output is correct?}~\cite{Sarkar2015}, \textit{How to
	combine models?}~\cite{Talbot2009}, \textit{Which model works
	best?}~\cite{Amershi2015}, or \textit{How do changes affect the rest of the
	system?}~\cite{Kulesza2008}. These and similar questions imply a focus on
recognising and handling system error, giving feedback, or analysing the system
to better work with it in the future.

\subsubsection{Interpretive Mindset}
An interpretive mindset \textit{strives to interpret system actions based on
	one's perception of and experience with the system and its output.} This mindset
embraces the notion of \textit{user experience}. 

When users adopt this mindset, they do not necessarily want to reach a
particular practical goal, but rather to understand the system based on a
certain experience they have made. For example, a social media user might want
to understand why posts of particular friends are not shown~\cite{Bucher2017}.
Moreover, an interpretive mindset might be adopted by users who do not
understand how they are being profiled, when they believe their feedback is not
being considered or feel they lack control over system output~\cite{Bucher2017}.

Examples for an (implied) interpretive mindset can be found in work on
transparency (e.g. \textit{How sensible -- and not arbitrary or random -- is the
	system?}~\cite{Rader2018}) and interpretability (e.g. \textit{What else can the
	model tell me about the world?}~\cite{Lipton2018}). Moreover, it is reflected in
many user questions articulated in work on scrutability, for example
\textit{What else does the system think I (don't) know?}, \textit{What would the
	system do if I did Y?}, or \textit{What does the system do for other
	people?}~\cite{Kay2013}. Although control also plays an important role in
research in this field, the underlying perspective is framed in terms of
experience with and perception of the system and its output rather than a
practical goal.

\subsubsection{Critical mindset}
A critical mindset \textit{stresses normative, ethical and legal reflection
	about intelligent systems.} 

This echoes the wider discussion about the \sysprop, such as
transparency, explainability and accountability (e.g.~\cite{Eiband2018,
	Hildebrandt2016, Mittelstadt2016, Wachter2017a, Wachter2017}). For example, a
user might critique a system's missing social
intelligence~\cite{Brown2019, Bucher2017} or might want to know why it was
developed in a certain way~\cite{Rader2018}. 
A critical mindset may thus be decoupled from system use.

Calls for support of critical inquiry are mainly found in work on system
accountability. User questions include \textit{How was the system developed and
	how is it continually being improved?}, \textit{What is informed by the
	intelligent system (i.e. reveal existence of intelligent decision-making and
	processing)?}, \textit{How fair and controllable is the
	system?}~\cite{Rader2018}, \textit{How fair is a system
	decision?}~\cite{Binns2018} or \textit{Can I trust this
	model?}~\cite{Ribeiro2016} and \textit{Should I trust this
	prediction?}~\cite{Herlocker2000,Ribeiro2016}.

\vspace{\baselineskip}
\noindent
\colorbox{highlight}{
    \begin{minipage}{.98\columnwidth} 
        \includegraphics[width=1em]{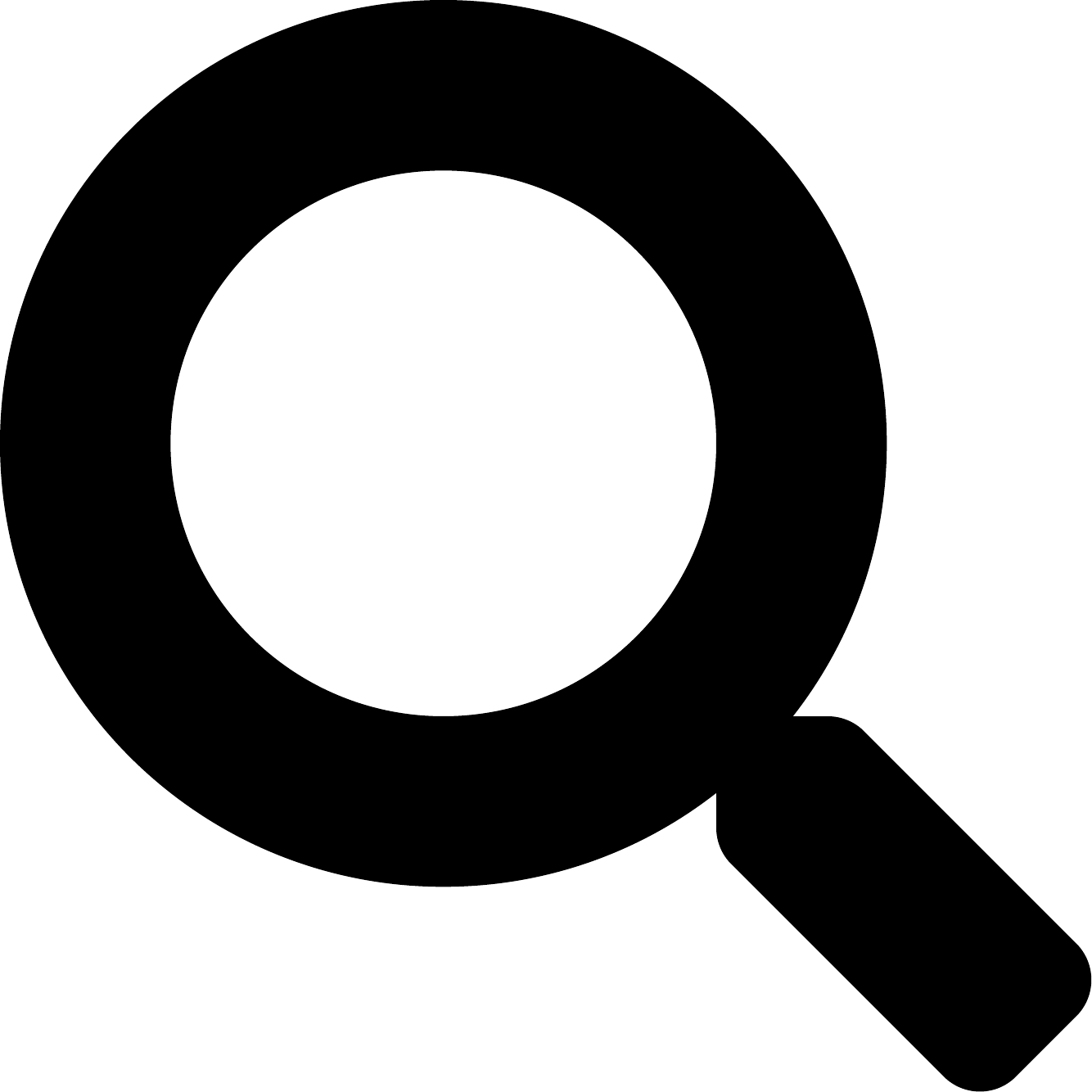}
        \textbf{User mindsets} help to understand what users seek to know when interacting with an intelligent system. We should thus make explicit which mindset(s) we assume as a basis for our work (e.g. utilitarian, interpretive or critical).
\end{minipage}
} 

%% file: sections/03b-involvement.tex
\subsection{How do Users Gain Knowledge? User Involvement}\label{sec:roles}

As introduced in the Scope and Foundations section of this paper, we assume that user understanding is built through the \textit{transmission of information} between user and system. 

Our analysis revealed that the great majority of work on the \sysprop{} envisions this transmission of information in the form of a \textit{dialogue}, that is as a ``cycle of communication acts channelled through input/output from the machine perspective, or perception/action from the human perspective''~\cite{Hornbaek2017}. Dialogue as an interaction concept inherently stresses the need for users to understand the system (and vice versa)~\cite{Hornbaek2017}.
It has even been argued that the characteristics of intelligent systems \textit{necessarily} involve some sort of dialogue in order for users to understand them~\cite{Sarkar2015}. Elements of dialogue, such as structuring interaction as \textit{stages}~\cite{Hornbaek2017}, are commonly found in work on the \sysprop. Most notably, end-user debugging~\cite{Kulesza2015} and interactive machine learning~\cite{Kapoor2010} make use of mixed-initiative interfaces~\cite{Horvitz1999}. It thus seems that the \sysprop{} almost \textit{imply} this concept of interaction, so closely are they interwoven with a dialogue structure. 
To support users in understanding intelligent systems, information may thus be transferred in two directions, either from user to system, or from system to user. From a user perspective, this determines \textit{how users gain knowledge} -- either through action (active) or perception (passive). 

Our second framework category, \textit{user involvement}, captures these two ways of gaining knowledge. \textit{User involvement} thus describes \textit{interaction possibilities to transfer information to or receive information from the system as a basis for user understanding}.
In the following sections, we distinguish work on the \sysprop{} according to these two directions. In general, a system might support involvement in multiple ways (e.g. via explanation, controls, visualisations) and thus imply transitioning between both directions during use, for example, through interactive visual explanations~\cite{Kulesza2015}.

\subsubsection{Active User Involvement (User-to-System)}
User questions such as \textit{How can I tell the system what I want?}~\cite{Kay2013}, \textit{How can I detect system errors?}~\cite{Kulesza2011}, or \textit{What do I have to change to correct the system?}~\cite{Kulesza2011} point to active users whose corrections and feedback are utilised by the system. The dialogue between system and user may be initiated by both sides and is then indeed based on turn-taking, as described earlier. For example, Alkan et al.~\cite{Alkan2019} presented a career goal recommender that literally employs a dialogue structure to suggest items and incorporate user feedback. Work on scrutability by Kay and Kummerfeld~\cite{Kay2013} emphasises the ``real effort'' users must make when inquiring into a system. Other work on the \sysprop, in particular in end-user-debugging and interactive machine learning, sees users in active roles as debuggers~\cite{Kulesza2010} or teachers~\cite{Amershi2014}. 

Systems support an active user by offering \textit{interface controls} tied to aspects of their ``intelligence'' (e.g. data processing, user model). These controls may enable users to experiment with the intelligent system (e.g. user model controls~\cite{Kay2013}). For example, if a user interface offers switches for certain data sources (e.g. in a settings view), users can actively experiment with the way that these data sources influence system output (e.g. switch off GPS to see how recommendations change in a city guide app; also see~\cite{Kay2013}). Moreover, Coppers et al.~\cite{Coppers2019} introduced widgets for system feedforward that allow for active inquiry to answer \textit{What if?} user questions such as \textit{What will happen if I click this checkbox?}. 
Another example is a separate, dedicated GUI for such experimentation, which allows users to directly set the values of certain system inputs and check the resulting output (e.g. ``intelligibility testing''~\cite{Lim2009a}).

Moreover, many visual explanations offer direct manipulation that also puts users into an active role: For instance, work in interactive machine learning proposed interactions with classifier confusion matrices to express desired changes in resulting decisions~\cite{Kapoor2010, Talbot2009}. Similarly, work on explanations for spam filtering enabled users to influence the classifier via interactive bar charts of word importance~\cite{Kulesza2015}.

\subsubsection{Passive User Involvement (System-to-User)}
User questions such as \textit{Why does the system think that I want/need X?}~\cite{Billsus1999}, \textit{Why did the system do X?}~\cite{Lim2009, Rader2018}, \textit{Why did it not do Y?}~\cite{Lim2009a} or \textit{How does the system produce an output?}~\cite{Rader2018} suggest that users want to get informed about the systems inner workings, but do not actively provide the system with feedback and corrections. Users may still initiate the dialogue with the system, but are then restricted to be recipients of information. This way of user involvement is typically assumed by work on transparency and explainability, where \textit{displaying information} about a system's inner workings is a common tool for user support. For example, related work proposed visual and textual explanations that show how recommendations are influenced by data from customers with similar preferences~\cite{Gedikli2014}. 
Further examples of supporting user understanding in a passive way include icons that indicate ``intelligent'' data processing~\cite{Eiband2018a}, interaction history~\cite{Hussein2010}, annotations for specific recommendations~\cite{Blanco2012}, and plots and image highlighting for classification decisions~\cite{Ribeiro2016} or recommendations~\cite{Tsai2019}.

\vspace{\baselineskip}
\noindent
\colorbox{highlight}{\begin{minipage}{.98\columnwidth} 
\includegraphics[width=1em]{figures/search-solid}
\textbf{User involvement} describes how user knowledge is built during interaction with a system. This depends on the direction of information transmission between user to system (e.g. users are involved in an active or passive way). We should explicitly state the nature of user involvement and how it is manifested in and supported through design.
\end{minipage}} 

%% file: sections/03c-outcomes.tex
\subsection{Which Knowledge Do Users Gain? Knowledge Outcomes}\label{sec:knowledge}
The envisioned result of the different \sysprop{} is knowledge that users gain about an intelligent system. However, this knowledge may refer to different aspects of the system and interaction, such as a specific recommendation~\cite{Pu2006} or the ``reasoning'' of a system~\cite{Ribeiro2016}.
To account for this variety of \textit{which knowledge users gain}, we introduce our third framework category, \textit{knowledge outcomes}. These \textit{characterise the nature of user understanding developed about an intelligent system}.
Overall, our analysis surfaced four different knowledge outcomes currently addressed in the literature (\textit{output}, \textit{process}, \textit{interaction}, and \textit{meta}). Since output and process knowledge are often confronted in the literature, we present them in one subsection here, too.

These knowledge outcomes are not unique to HCI or intelligent systems. For example, output and process knowledge can be found in work on theory on gaining knowledge in practice~\cite{Lynham2002}. Moreover, work on complex problem solving articulates output, process and structural know\-ledge~\cite{Schoppek2002}, the latter being similar to our interaction knowledge.

We also introduce two \textit{qualities} of  knowledge emerging from the reviewed literature. Borrowing established terms for knowledge qualities in applied research theory~\cite{Hevner2004, Lynham2002}, we summarise them as \textit{rigour} and \textit{relevance} of know\-ledge.

\subsubsection{Output and Process Knowledge}
\textit{Output knowledge} targets \textit{individual instances} of an intelligent system (e.g. understanding a specific movie recommendation). 
In contrast, \textit{\knowprocess{}} targets the \textit{system's underlying model and reasoning steps} (e.g. the workings of a neural network that processes movie watching behaviour).

Explainability research in particular distinguishes between explanations for instances and models. For example, Ribeiro et al.~\cite{Ribeiro2016} explain classifiers with regard to two questions, \textit{Should I trust this prediction?} and \textit{Can I trust this model?}. Therefore, they design for both output and process knowledge. These two knowledge types also motivate the \textit{what} and \textit{how} questions posed by Lim and Dey in their work on intelligibility~\cite{Lim2009} (e.g. \textit{What did the system do?}). Also, Rana and Bridge~\cite{Rana2018} introduced \textit{chained} explanations (called ``Recommendation-by-Explanation'') to explain a specific output to users. Moreover, work on accountability makes system reasoning accessible to users to support the development of \knowprocess~\cite{Binns2018}.

\subsubsection{Interaction Knowledge}
Our knowledge type \textit{\knowinteract{}} describes knowing how to do something in an interactive intelligent system. For example, supporting users in gaining this type of knowledge motivates questions in work on scrutability (e.g. \textit{How can I tell the system what I want to know (or not)?}~\cite{Kay2013}), interactive machine learning (e.g. \textit{How to experiment with model inputs?}~\cite{Amershi2014}), and end-user debugging (e.g. \textit{How can I tell the system why it was wrong?}~\cite{Kulesza2011}, \textit{How can I correct system errors?}~\cite{Kulesza2010}).

\subsubsection{Meta Knowledge}
\textit{Meta knowledge} captures system-related knowledge \textit{beyond} interaction situations, such as information from a developer blog. For example, meta knowledge motivates some questions in work on transparency, such as \textit{How is the system developed and how is it continually improved?} by Rader et al.~\cite{Rader2018}. They also explicitly add \textit{Objective} explanations that inform users about how ``a system comes into being'' that result in \knowmeta{} (e.g. development practices and contexts).
Moreover, this knowledge type is a main driver of work on accountability, in which computer science overlaps with journalism: For instance, Diakopoulus ``seeks to articulate the power structures, biases, and influences'' of intelligent systems~\cite{Diakopoulos2015}.

\subsubsection{Rigour and Relevance of Knowledge}

\textit{Rigour:} Kulesza et al. propose the concepts of \textit{soundness} and \textit{completeness} in their work on explanations in intelligent systems~\cite{Kulesza2013, Kulesza2015} -- \textit{soundness} is truthful explanation, and \textit{completeness} means explaining the whole system. 
Gilpin et al.~\cite{Gilpin2018} also refer to completeness, yet understand it as supporting anticipation of system behaviour in more situations.
For an overarching view, we generalise this to a broader level: We regard soundness and completeness as facets of \textit{rigour}. Linked back to the work by Kulesza et al.~\cite{Kulesza2015, Kulesza2013}, this means that a rigorous explanation, and the resulting understanding of a system, should be sound and complete. 

\textit{Relevance:}
A rigorous understanding does not need to be useful. We argue that this aspect should be of explicit interest for a pragmatic HCI perspective.
We thus consider \textit{relevance} as another general quality of knowledge~\cite{Lynham2002} that is crucial to make explicit in the specific context of user understanding of intelligent systems. 
This quality highlights our pragmatic view: Elements like explanations are valuable if they add utility, that is, if they help users to gain knowledge that is relevant for \textit{using} the system in better ways and towards better outcomes. 
In this pragmatic sense, this quality echoes Kulesza et al.'s suggestion to ``not overwhelm'' users with (irrelevant) information~\cite{Kulesza2015, Kulesza2013}. What is relevant to know, and to which extent, may also depend on factors such as task and complexity of the system~\cite{Bunt2012}.

\vspace{\baselineskip}
\noindent
\colorbox{highlight}{\begin{minipage}{.98\columnwidth} 
\includegraphics[width=1em]{figures/search-solid}
\textbf{Knowledge outcomes and qualities} characterise and make explicit what kind of user understanding a system seeks to facilitate. We should thus articulate the goals of our work (e.g. output, process, interaction, and meta knowledge) and reflect on rigour and relevance of the respective kind of knowledge.
\end{minipage}}

%% file: sections/04-discussion.tex
\section{Framework Application: Structuring Past \& Future Work}\label{sec:discussion}

We have presented three categories for supporting user understanding of intelligent systems as emerged from our analysis of the literature -- \textit{user mindsets}, \textit{user involvement}, and \textit{knowledge outcomes}. These categories highlight differences and commonalities between work on the \sysprop{} and serve as a conceptual framework of supporting users in understanding intelligent systems. 

Our framework introduces an overarching user-centric structure to the field that abstracts from the fractured terminological landscape. 
We now propose to use our framework categories as a means for researchers in HCI and adjoining fields to clarify and make explicit the assumptions of their work, and to structure past and future work and discussions about how to support users in understanding intelligent systems. The boxes presented throughout this article provide inspiration on what to consider. The following sections highlight further applications of our framework.

\subsection{Structuring Past Approaches and Solution Principles}

\begin{figure*}[t]
    \centering
    \includegraphics[width=0.85\textwidth]{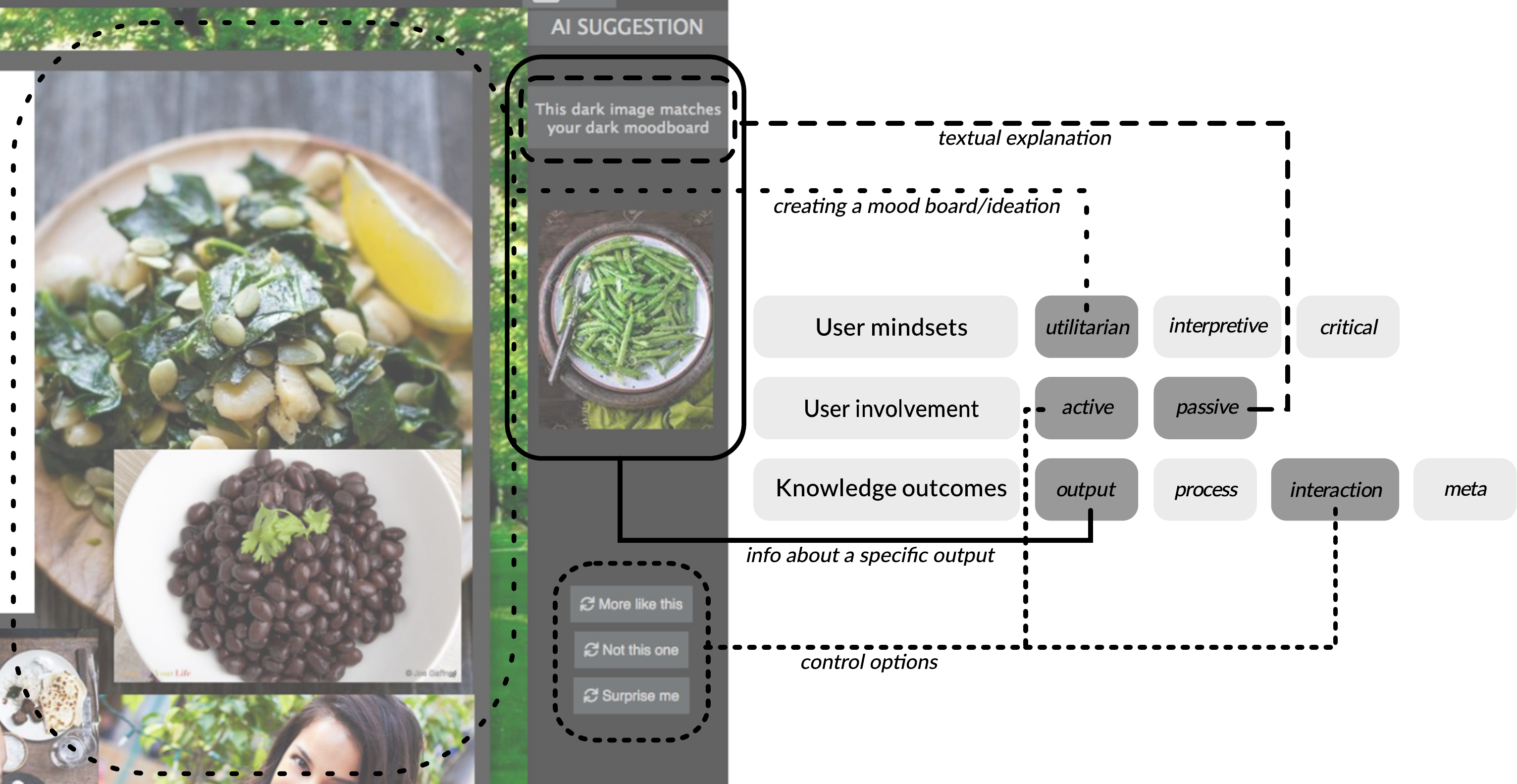}
     \caption{Application of our framework using Koch et al.'s system-aided mood board creation tool~\cite{Koch2019} as an example. System image from Koch et al.~\cite{Koch2019} used with the authors' permission.}
     \label{fig:application}
\end{figure*}

In Figure~\ref{fig:application}, we demonstrate the application of our framework based on an interactive intelligent mood board creation tool by Koch et al.~\cite{Koch2019} as an example for system-aided design ideation. In their prototype, system and user collaborate to find suitable imagery, the system making suggestions which the user can accept or discard. For each suggestion, the system offers a textual explanation and different feedback options. In terms of \textit{user involvement}, this approach thus supports both an \textit{active} user (via options for feedback and correction and turn-taking), as well as a \textit{passive} one (via information about why an image was suggested). With regard to \textit{knowledge outcomes}, users gain \textit{relevant} knowledge about a specific \textit{output} (a specific image suggestion) and \textit{interaction} knowledge about how to control future system suggestions (by telling the system what they like or not). Overall, the prototype is designed so as to best support users in the creation of a mood board as an ideation activity, and thus implies a \textit{utilitarian} mindset.  

Moreover, Table~\ref{tab:solution_principles} presents other exemplary solution principles from the literature to illustrate how user mindsets, user involvement and knowledge outcomes may be used to structure past work. We do not claim to provide a comprehensive survey in our work, but selected these examples to show the diversity of approaches in our paper set. 

Studying the approaches as arranged in Table~\ref{tab:solution_principles} reveals interesting structures: 
For example, explanations appear across the charted space and thus could be seen as the go-to building block for many solution principles. They commonly provide output and process knowledge via text and/or plots~\cite{Gedikli2014, Hussein2010, Kizilcec2016, Kulesza2011, Kulesza2015, Lim2011, Ribeiro2016}. 
Sometimes these representations also allow for interactivity and user corrections~\cite{Kulesza2011, Kulesza2015, Lim2009a}, in particular when explanations are referred to in work on scrutability, end-user debugging, and interactive machine learning~\cite{Kapoor2010, Kay2013, DeRussis2018, Talbot2009}. Explanations commonly arise from utilitarian mindsets, yet they also appear in work with interpretive and critical questions~\cite{Blanco2012, Diakopoulos2016, Kay2013, Rader2018}.

\subsection{Reflecting on Your Own Approach}

\subsubsection{Reframing User Questions}
User questions are a helpful way to uncover users' information needs. Our framework can be used to re-frame such questions in related work, in particular by reconsidering the underlying mindsets to view questions from a novel angle:
For example, a question such as \textit{Why did the system do X?} is currently mostly tied to a context implying a utilitarian mindset~\cite{Lim2009}. However, this question could also reflect other mindsets and thus different underlying user motives for inquiry, depending on the envisioned context. Design solutions to this question could then foster utilitarian (e.g. explain feature influences), interpretive (e.g. explain in terms of a user's daily life context), critical (e.g. explain system decision given a community's norms), or all three mindsets.

\begin{table*}
\newcolumntype{L}{>{\raggedright\arraybackslash}X}
\renewcommand{\arraystretch}{1.5}
\tiny
\centering
\scriptsize
\resizebox{0.925\textwidth}{!}{%
\begin{tabularx}{\textwidth}{@{}ccLLLLcc@{}}
\multicolumn{1}{l}{} & \multicolumn{1}{l}{} & \multicolumn{4}{c}{\mytabhead{Knowledge Outcomes}} & \multicolumn{1}{l}{} & \multicolumn{1}{l}{} \\
\multicolumn{1}{l}{} & \multicolumn{1}{l}{} & \multicolumn{1}{c}{\mytabheads{Output}} & \multicolumn{1}{c}{\mytabheads{Process}} & \multicolumn{1}{c}{\mytabheads{Interaction}} & \multicolumn{1}{c}{\mytabheads{Meta}} & \multicolumn{1}{l}{} & \multicolumn{1}{l}{} \\
\cmidrule{3-6}
\multirow{6}{*}[-6cm]{\mytabhead{\rot{User Involvement}}} & \multirow{1}{*}[-.5cm]{\mytabheads{\rot{Active}}} & Visualisations explain classification via features, user ``explains back'' corrections by manipulating these plots~\cite{Kulesza2011, Kulesza2015}. \newexample
``Reasons'' tabs show sensor-specific visualisations that explain current prediction~\cite{Lim2011}. & Users build if-rules in an editor, system detects problems via simulation (e.g. loops), user corrects them~\cite{DeRussis2018}. \newexample
``Experimentation view'' allows users to try out inputs and see system output~\cite{Lim2009a}. \newexample
Confusion matrices show current state of classifier~\cite{Kapoor2010, Talbot2009}. & Natural language dialogue enables users to ask system what they could do next~\cite{Tintarev2014}. \newexample
Users manipulate confusion matrix to change model~\cite{Kapoor2010}, including re-combining multiple models~\cite{Talbot2009}. & Beyond runtime: Open source enables code audits and facilitates understanding of system~\cite{Burrell2016}. \newexample & \multirow{2}{*}[-1.5cm]{\mytabheads{\rot{Utilitarian}}} & \multirow{6}{*}[-6cm]{\mytabhead{\rot{Mindset}}} \\
& \multirow{1}{*}[-.5cm]{\mytabheads{\rot{Passive}}} & \cellcolor[HTML]{EFEFEF} Bar charts and image regions show importance of predictors to explain specific classification~\cite{Ribeiro2016}. \newexample Visualisations show similar users' input to explain recommendation~\cite{Gedikli2014}. \newexample Tree of images of training instances explains classification~\cite{Yang2020}.
 & \cellcolor[HTML]{EFEFEF} Explaining a classifier by explaining multiple specific classifications~\cite{Ribeiro2016}. \newexample
Animation shows learning algorithm at work~\cite{Jackson1997}. \newexample Text highlighting shows which words contributed to meeting detection in emails.~\cite{Kocielnik2019} & \cellcolor[HTML]{EFEFEF} List shows user's past interactions to explain specific recommendation~\cite{Hussein2010}. \newexample Step-by-step explanations of trigger-action rules by simulating user and system actions~\cite{Corno2019}. & \cellcolor[HTML]{EFEFEF} Beyond single systems: Educate public in computational skills to facilitate system understanding overall~\cite{Burrell2016}. & & \\
\cmidrule{3-6}
& \multirow{1}{*}[-.5cm]{\mytabheads{\rot{Active}}} & Rule-based reasoning system verbalises system decisions in natural language dialogue with user~\cite{Tintarev2014}. \newexample
 & Separate profile page displays current user model~\cite{Kay2013}. \newexample
Natural language dialogue enables users to ask system why it has not decided differently.~\cite{Tintarev2014} \newexample
Constructivist learning: user manipulates system and updates mental model of it based on resulting changes in output~\cite{Sarkar2016}. & Separate profile page enables users to edit what the system knows about them~\cite{Kay2013} \newexample
``Algorithmic profiling management'': Profile page reveals what system knows and how this influences content, including past interactions; controls enable modifications~\cite{Alvarado2018}. 
& ``Algorithmic UX'' beyond interaction: Users engage in communication and relationship building with intelligent agents~\cite{Oh2017}. & \multirow{2}{*}[-1.5cm]{\mytabheads{\rot{Interpretive}}} & \\
& \multirow{1}{*}[-.5cm]{\mytabheads{\rot{Passive}}} & \cellcolor[HTML]{EFEFEF} Icon indicates which system output is influenced by AI~\cite{Eiband2018a}. \newexample Text analysis extracts relevant sentences from reviews to show along with recommendation~\cite{Donkers2020}.
& \cellcolor[HTML]{EFEFEF} Output shown with textual explanation of the decision process~\cite{Kizilcec2016}. \newexample
Animation indicates how system output is generated (e.g. dice roll for randomness)~\cite{Eiband2018a}. & \cellcolor[HTML]{EFEFEF} Explain Recommendations with usage statistics (e.g. global popularity, repeated interest)~\cite{Blanco2012}. & \cellcolor[HTML]{EFEFEF} Beyond code: Educate/sensitise developers and decision makers to consequences of systems~\cite{Burrell2016}. \newexample
Icons indicate when and for which high-level goal (e.g. ads) user data is processed by the system~\cite{Siljee2015}. & & \\
\cmidrule{3-6}
& \multirow{1}{*}[-.5cm]{\mytabheads{\rot{Active}}} & ``Algorithmic accountability reporting'': Journalists report on black box systems, e.g. by trying out inputs systematically~\cite{Diakopoulos2015, Diakopoulos2017}. & Beyond system understanding: Society must look not into systems but across them, that is, see their role within a larger network of actors (incl. humans and institutions)~\cite{Ananny2018}. & Challenging the system: People learn from past interactions and output how to challenge system intelligence and its normative implications through unexpected or malicious input (e.g. manipulating public chatbot via twitter)~\cite{Neff2016}. & Beyond system use: People discuss and reflect on social implications and context of the system's output~\cite{Brown2019, Schlesinger2018, Veale2018}. & \multirow{2}{*}[-1.5cm]{\mytabheads{\rot{Crictical}}} & \\
& \multirow{1}{*}[-.5cm]{\mytabheads{\rot{Passive}}} & \cellcolor[HTML]{EFEFEF} Record models, algorithms, data, decisions for later audit~\cite{ACM2017}. \newexample
Annotate recommended content pieces with indicators for quality/reliability of their source (e.g. for news)~\cite{Lazer2018}. & \cellcolor[HTML]{EFEFEF} Textual explanations of system intelligence on a high level, not integrated into the system (e.g. articles about system)~\cite{Rader2018}. \newexample
Explaining the logic behind an algorithm with another algorithm~\cite{Brkan2017}. & \cellcolor[HTML]{EFEFEF}  ``Algorithmic Imaginary'': People develop understanding of intelligent systems and how to influence them based on how ``they are being articulated, experienced and contested in the public domain''~\cite{Bucher2017}. & \cellcolor[HTML]{EFEFEF} Textual explanations of developers' intentions on a high level, not integrated into the system (e.g. articles about system)~\cite{Rader2018}. & & \\
\cmidrule{3-6}    
\end{tabularx}
}
\caption{Examples of approaches and solution principles for supporting user understanding of intelligent systems, structured through our framework. This is not a comprehensive survey; examples were selected to illustrate the diversity of approaches in the literature.}
\label{tab:solution_principles}
\end{table*}

\subsubsection{Explicating Perspectives with Mixed Mindsets}
Related, reflecting on the three mindsets presented here can help to discover structure in perspectives that mix multiple mindsets: For example, a recent set of 18 guidelines for interaction with AI~\cite{Amershi2019} contained mostly utilitarian guidelines (e.g. ``Support efficient dismissial'', ``Provide global controls''), yet two stand out as following a critical mindset (``Match relevant social norms'', ``Mitigate social biases''). Our lens allows to clarify and explicate this mix, revealing, in this example, that the guidelines already follow a broader perspective than the related work itself alluded to with its stated focus ``on AI design guidelines that [...] could be easily evaluated by inspection of a system's interface''. Such analysis could help to structure discussions about similarly mixed perspectives.

\subsubsection{Explicitly Determining Relevance}
What is relevant to know for users is considered differently across the \sysprop. Highlighting relevance and rigour (see section on \textit{knowledge outcomes} and Table~\ref{tab:system_qualities_questions} last column) thus helps to reflect on what we consider important, for whom, and to what extent -- and how we choose to determine it. 

For example, explanation design often involves non-expert users, possibly via a user-centred design process~\cite{Eiband2018a} or scenario-based elicitation~\cite{Lim2009a} to inform what is relevant, what should be explained, and to what extent.

In contrast, interactive machine learning focuses on information that is relevant to the task of the \textit{system} (e.g. to improve a classifier), which is often operated by experts~\cite{Amershi2014}. Therefore, what is relevant here (and to what extent) is foremost informed by the machine learning task, and less so by studying or asking end-users. This can be seen, for example, in the UI elements derived in a recent survey on interactive machine learning~\cite{Dudley2018}, which are closely coupled to the machine learning task (e.g. they serve ``sample review'', ``feedback assignment'', etc.). 

As another example, work on end-user debugging presents an action-focused middle-ground, between user-focused (as explainability) and system-focused (as interactive machine learning): Here, resulting knowledge should help users to \textit{make} the system more relevant to them, for example, by correcting system errors from the users' point of view; users may be both experts~\cite{Amershi2015} or non-experts~\cite{Kulesza2011}.

\subsubsection{Informing Methodology}
Our framework may be used to motivate methodological choices, for example when informing or evaluating the design of a new approach for supporting user understanding: 

For instance, work catering to a utilitarian mindset might benefit from a controlled environment and precise measurements in a lab study. Even simulation of system decisions might be a (first) option~\cite{Doshi-Velez2017}. A lab study might also be a suitable choice to evaluate support for developing interaction knowledge, since users can be directly observed during interaction (e.g. see~\cite{Kay2013, Talbot2009}).
In contrast, if a design or research question targets an interpretive mindset and/or meta knowledge, it might be worthwhile or required to study user and system in their daily contexts of use (e.g. see lab vs field study in education context in~\cite{Kay2013}). 
The same holds for work motivated by a critical mindset, yet other methods exist here as well, such as online surveys or data analyses of views expressed through mass media and social network discussions~\cite{Lazer2018}, or policy and legal texts~\cite{ACM2017, GDPR, Goodman2016}.

\subsection{Going Beyond}

\subsubsection{Inspiring New Approaches in Research and Practice}\label{sec:inspiring}
Our framework may provide inspiration for new approaches and solution principles. While the target audience of this paper are foremost researchers in the field, we believe that the framework might also be used by practitioners, for example, as a brainstorming tool for prototype development. 

We illustrate this using the proposed mindsets: UIs could support users in examining the system with different mindsets via ``modes'' for explanation views. Users could then switch between utilitarian explanations (e.g. explain a recommendation with product features) and interpretive or critical ones (e.g. explain system beliefs about a user, such as that the user is part of a certain target group; reveal that recommendations are assembled by an AI and not by a human, cf.~\cite{Eiband2018a}). 

A more radical solution could offer three different views of the system that display or hide UI elements depending on the respective ``mode''. Or the mindsets might simply help to decide which user approach to support in a system, and to identify those remaining unaddressed so far.
Prototypes could then be tested with the respective end-users of the application. Yet, the actual generative power of our framework has to be validated in the future.

\subsubsection{Understanding Systems Beyond Interaction Situations}
The critical mindset and meta knowledge capture a crucial difference between traditional (non-intelligent) systems and what we see today and what is yet to come: Systems are increasingly interwoven with our lives, be it in everyday applications or in areas of consequential decision-making (e.g. financial, medical or legal). Their effects thus do not remain limited to a particular interaction situation. It is important that we as researchers reflect on the impact of the systems we design beyond the duration of direct use. This also includes reflections on when and how intelligent systems can learn compared to humans in the same roles~\cite{Alkhatib2019}. 
Examples for work in such a larger context are presented in Table~\ref{tab:solution_principles}, in particular in the \textit{critical} and \textit{meta} areas (e.g.~\cite{Ananny2018, Burrell2016, Diakopoulos2015, Diakopoulos2017}). Connections with HCI in such work commonly refer to \textit{accountability} and \textit{transparency} of intelligent systems. 

\subsubsection{Motivating Connections Beyond HCI \& Machine Learning/AI}
We see recent calls for more joint research at the intersection of HCI and AI to improve system understanding~\cite{Abdul2018}. However, this is mostly motivated by utilitarian or interpretive mindsets. Thus, another related key takeaway is to draw attention to interdisciplinary connections via the \textit{critical} of the three mindsets proposed in this article: As is evident from recent ``AI and data scandals'' (e.g.~\cite{Grassegger2017web, Lazer2018, Neff2016}), developing more understandable (and accountable) intelligent systems also needs to be addressed in a wider view (cf. third wave HCI~\cite{Bodker2015}), for example across networks of human and AI actors~\cite{Ananny2018}.
More generally, fruitful connections could span considerations from fields like journalism~\cite{Diakopoulos2015, Diakopoulos2017} and communication~\cite{Lazer2018, Neff2016}, policy~\cite{Alkhatib2019}, sociology~\cite{Ananny2018} and education~\cite{Burrell2016}, and ethical and legal concerns~\cite{Brkan2017, Eiband2018}.

%% file: sections/05-conclusion.tex
\section{Conclusion}

Intelligent systems tend to violate UI principles, such as predictable output~\cite{Amershi2014, Dudley2018}, which makes them difficult to understand and use. 
To address this, researchers, practitioners, policy-makers and the general public call for \textit{\sysprop{}} such as transparency, scrutability, explainability, interpretability, interactivity, and so on. 
However, these terms are often blurred and employed with varying interpretations. This impedes conceptual clarity of the very properties that are envisioned to foster users' understanding of intelligent systems. 

This review responds to this lack of conceptual clarity with an analysis and discuhssion of theoretical concepts and prototype solutions from the literature: We make explicit the diversity of different implied views on \textit{user mindsets}, \textit{user involvement}, and \textit{knowledge outcomes}. 

In conclusion, we provide researchers with a framework to (1) clearly motivate and frame their work, (2) draw connections across work on different \sysprop{} and related design solutions, and (3) articulate explicitly their underlying assumptions and goals.
With our work, we thus hope to facilitate, structure and advance further discussions and research on supporting users' understanding of intelligent systems.